\documentclass[twocolumn,showpacs,preprintnumbers,amsmath,amssymb,superscriptaddress,longbibliography,pre,reprint]{revtex4-1}
\usepackage{color}

\usepackage{graphicx}
\usepackage{dcolumn}
\usepackage{bm}
\usepackage{subfigure}
\usepackage{amssymb}
\usepackage{tabularx}
\usepackage{amsmath}
\usepackage{ulem}
\usepackage{xspace}

\renewcommand{\vec}{\mathbf}
\renewcommand{\emph}{\textit}

\newcommand{\Tpar}{\ensuremath{T_\parallel}\xspace}
\newcommand{\Tperp}{\ensuremath{T_\perp}\xspace}

\newcommand{\apar}{\ensuremath{\alpha_\parallel}\xspace}
\newcommand{\aperp}{\ensuremath{\alpha_\perp}\xspace}
\begin{document}

\title{Spontaneous generation of a temperature anisotropy in a strongly coupled magnetized plasma}

 \author{T. Ott}
 \affiliation{%
     Christian-Albrechts-University Kiel, Institute for Theoretical Physics and Astrophysics, Leibnizstra\ss{}e 15, 24098 Kiel, Germany
 }%
 \author{M. Bonitz}%
 \affiliation{%
     Christian-Albrechts-University Kiel, Institute for Theoretical Physics and Astrophysics, Leibnizstra\ss{}e 15, 24098 Kiel, Germany
 }%
  \author{P.~Hartmann}
\affiliation{
Institute for Solid State Physics and Optics, Wigner Research Centre for Physics, H-1525 Budapest, P.O.B 49, Hungary}
 \author{Z.~Donk\'o}
\affiliation{
Institute for Solid State Physics and Optics, Wigner Research Centre for Physics, H-1525 Budapest, P.O.B 49, Hungary}

\date{\today}

\begin{abstract}
 A magnetic field was recently shown to enhance field-parallel heat conduction in a strongly correlated plasma whereas cross-field conduction is reduced. Here we show that in such plasmas, 
 the magnetic field has the additional effect of inhibiting the isotropization process between field-parallel and cross-field temperature components thus leading 
 to the emergence of strong and long-lived temperature anisotropies when the plasma is locally perturbed. An extended heat equation is shown to describe this process accurately. 
\end{abstract}

  \maketitle

\section{Introduction}

Heat conduction in strongly coupled plasmas (SCPs)---i.e., plasmas in which the potential energy exceeds the kinetic energy---is a crucial 
issue for many physical phenomena and experimental challenges. Experiments in which strong coupling is realized include 
dusty plasmas~\cite{Ivlev2012,Bonitz2010a}, trapped ions~\cite{Jensen2005}, and ultracold plasmas~\cite{Killian2007}. Such conditions also occur naturally, e.g., in white dwarf stars or 
the outer layers of neutron stars~\cite{Shapiro1983, Potekhin2010}. 

Frequently, these plasmas are subject to strong magnetic field which strongly modifies their ability to conduct heat along and across the field lines. 
Such situations occur, e.g., in magnetars~\cite{Chugunov2007,Potekhin2015} or in laser fusion experiments at NIF and Omega through self-generated 
magnetic fields~\cite{Rygg2008}. In magnetized liner inertial fusion experiments at Sandia~\cite{Gomez2014}, very strong magnetic fields 
are generated during the compression phase. An easily accessible experiment for strong coupling and strong ``quasi''-magnetization is provided 
by rotating dusty plasmas which were recently demonstrated to accurately mimic the physics of magnetized SCPs~\cite{Kahlert2012,Hartmann2013,Bonitz2013}. 

The effect of magnetic fields on the heat conduction in SCPs was recently calculated from equilibrium molecular dynamics simulations~\cite{Ott2015a, Ott2016}. 
It was found that the interplay of strong correlation and strong magnetization leads to the counter-intuitive result that the heat conductivity 
is \emph{enhanced} along the field lines, in contrast to the situation in weakly correlated plasmas~\cite{Braginskii1965}. Across the 
field lines, heat conduction is reduced by the magnetic field towards a residual minimum value which is due to phonon-like energy transfer~\cite{Ott2015a}. 

In this work, we report on an extension of these investigations in which the energy dissipation from a localized perturbation is explored. The key 
difference here is that the assumption of local thermal equilibrium is relaxed to the more general case of anisotropic temperatures, i.e., 
different temperatures \Tpar and \Tperp along and across the magnetic field.

The formation of such a temperature anisotropy in a magnetic field is well-known from weakly collisional plasmas. A prominent example is the solar wind, in which 
the proton temperatures \Tpar and \Tperp depend on the orientation relative to the magnetic field~\cite{Marsch1982}. The effect is limited to a 
temperature ratio of about ten by micro-instabilities such as the mirror and firehose instability~\cite{Schlickeiser2011}.
Critically, in such a plasma, the anisotropy is \emph{driven by anisotropic heating and cooling mechanisms}, e.g. the Chew-Goldberger-Low double-adiabatic
expansion~\cite{Chew1956} or cyclotron-resonant absorption of Alfv\'en waves~\cite{Maruca2011}. 

Conversely, we consider an \emph{isotropic temperature perturbation} of a strongly coupled plasma in a magnetic field, i.e., there is no externally imposed 
anisotropy and the formation of the temperature difference along and across the magnetic field follows from the intrinsic properties of the magnetized plasma. As 
we will show, this effect is generated from an interplay of anisotropic heat conduction and the efficient suppression of 
isotropization due to the preservation of the cyclotron energy in a magnetic field, i.e., the O'Neil many-body invariant~\cite{O’Neil1983,O’Neil1985}. 

The remainder of this work is organized as follows: First, in Sec.~\ref{sec:model}, we give details of our model and the simulations. In Sec.~\ref{sec:iso}, 
we consider the isotropization process in a magnetized SCPs. We will show that 
the cyclotron energy is almost conserved and test Dubin's prediction for the magnetic-field dependence of the isotropization timescale~\cite{Dubin2008}. 
In Sec.~\ref{sec:temprelax}, we consider the relaxation of an \emph{unmagnetized} SCP from a local temperature perturbation and compare with the results 
of equilibrium simulations. In Sec.~\ref{sec:magtemprelax}, we then consider the situation in a magnetized SCP in which both the isotropization and the 
relaxation time scale are influenced by the magnetic field and a temperature anisotropy is formed from an isotropic temperature inhomogeneity. We 
conclude in Sec.~\ref{sec:summary}.

\section{Model and Simulation}
\label{sec:model}

We adopt the paradigmatic one-component model of a three-dimensional plasma in which a single particle species of uniform 
mass $m$ and charge $q$ is considered against the background of a weakly polarizable delocalized second species. The 
interaction of the particles is then given by a screened Coulomb potential, i.e. a Yukawa- or Debye-H\"uckel-potential, 
\begin{align}
 \phi(r)=\frac{q^2}{r} \exp{(-{\tilde \kappa} r)}, 
\end{align}
where the Debye length $\tilde \kappa^{-1}$ is a parameter that characterizes the range of the potential. We choose a fixed value of $\kappa = \tilde\kappa a = 2$, where
$a=\left[3/\left (4 n \pi \right) \right]^{1/3}$ is the Wigner-Seitz radius. 
The equilibrium temperature is fixed by choosing a value for the Coulomb coupling parameter, $\Gamma= q^2/(ak_BT)=100$. 

In order to model isotropization and the temperature perturbation, we initially couple the system to a heat bath so the equations of motion are 
given by Langevin equations ($i=1\dots N$),
\begin{align}
m \ddot{{\vec r}}_i =\vec F_i +m \omega_c \dot{\vec r}_i \times \hat{\vec e}_B - m \nu^b \dot{\vec r}_i + \vec R_i(\xi, t)  \label{eq:langevin} 
\end{align}
where $\vec F_i$ is the force on the particle located at $\vec r_i$ due to all other particles, $\omega_c=qB/(mc)$ is the cyclotron frequency,  $\hat{\vec e}_B$ 
is a unit vector along the magnetic field, and $\nu^b$ is the heat bath coupling frequency. 
The term $\vec R_i(\xi,t)$ is a Gaussian white noise with zero mean and standard deviation
\begin{align}
   \langle R_{\alpha,i}(\xi,t_0)R_{\beta,j}(\xi,t_0+t)\rangle=2k_BT_\alpha(\xi)m\nu^b\delta_{ij}\delta_{\alpha\beta}\,\delta(t). \label{eq:langevin2} 
\end{align} 
Here, $\alpha,\beta\in \{x,y,z\}$ are Cartesian coordinates. 
The particular Cartesian coordinate $\xi$ is the direction along which the temperature profile is allowed to vary spatially. Note
that the temperature profile $T_\alpha(\xi)$ can take different values for different temperature components~$\alpha$. 

We solve the equations of motion using molecular dynamics simulation~\cite{Chin2008} 
for $N=265\,326$ particles situated in a cube of side length $L= (4\pi N/3)^{1/3} a = 103.6 a$ with periodic boundary conditions. In 
the following, we consider the strength of the magnetic field as the ratio $\beta=\omega_c/\omega_p$ of the cyclotron frequency  
and the plasma frequency $\omega_p= [ 4\pi q^2 n/m]^{1/2}$. The latter is also used to normalize times and other frequencies, as well.

\section{Isotropization}
\label{sec:iso}

The isotropization in a plasma with different temperatures \Tpar and \Tperp along and across a magnetic field 
is described by~\cite{Huba2009}
 \begin{align}
  \frac{dT_\perp}{dt} &= -\nu (T_\perp - T_\parallel),\\[-0.2ex]
  \frac{dT_\parallel}{dt} &= 2\nu (T_\perp - T_\parallel),
 \end{align}
where $\nu$ is the isotropization rate. 

For a magnetized one-component plasma, O'Neil and Hjorth showed that the cyclotron energy is an adiabatic many-body invariant~\cite{O’Neil1983,O’Neil1985}, i.e., an 
``almost conserved'' quantity. This in turn implies that the isotropization rate becomes arbitrarily small as the magnetic field increases and 
temperature anisotropies are increasingly long-lived. 

Dubin and Anderegg \emph{et al.} explicitly calculated and measured the isotropization rate for weakly and 
strongly coupled plasmas~\cite{Dubin2005,Dubin2008,Anderegg2009,Anderegg2010}. 
The validity regimes of these calculations depend (via the the mean adiabaticity $\kappa^\ast=\sqrt{6}\beta\Gamma^{3/2}$) on 
the ratio of $\Gamma$ and $\beta$ (which are both dimensionless):
For $\Gamma/\beta < \sqrt{6}$ and $\Gamma<1$ (``weak screening``), the isotropization frequency is $\nu\sim\exp(-\beta^{2/5})$. 
For $\Gamma/\beta < \sqrt{6}$ and $\Gamma>1$ (''strong screening``), the isotropization scales as before but is enhanced by a factor $g(\Gamma)$ (Salpeter enhancement). 
Finally, for $\Gamma/\beta > \sqrt{6}$ (''pycnonuclear regime``), there is no definite theory for the scaling of the isotropization frequency.

It is known that the relevant magnetic field strengths at which a strongly coupled plasma ($\Gamma=100$) is appreciably influenced by the 
magnetic field is on the order of unity, $\beta\approx 1$~\cite{Dzhumagulova2014,Ott2013,Ott2012,Hou2009d,Bonitz2010,Ott2011,Ott2013a,Bernu1981,Ott2011c,Feng2014,Ott2014}. 
This indicates that the system is in the pycnonuclear regime and our simulations can provide novel insight into the physics of SCPs.

Intuitively, the isotropization rate can be measured by first coupling the two temperatures \Tpar and \Tperp to separate heat baths as per Eq.~\eqref{eq:langevin} 
and then monitoring equipartition after removing the heat bath coupling. However, in the strongly coupled plasma regime, a large fraction of the energy difference 
can be absorbed into structural rearrangement, i.e., potential energy, rendering such straightforward measurement impossible. Instead, we make use 
of an indirect measurement approach which relies on the competition between an (artificial) heat bath anisotropy and the (physical) isotropization process in plasmas. 

In this method, the field-parallel and cross-field components are coupled to two independent heat baths whose temperatures differ. 
The coupling frequency $\nu^\textrm{b}$ to the heat baths is slowly increased and the response of the system temperature measured. 
The magnetic field is oriented along the $z$-axis, so that $T_\parallel=T_z$ and $T_\perp = T_{xy}$.

The temperatures of the two heat baths are chosen so that their ratio $A=T^b_z/T^b_{xy}=1.2$ corresponds to an anisotropy of 20\%, where the superscript 'b' indicates bath temperatures. 
At a given bath coupling frequency $\nu^\textrm{b}$, the system temperatures are stationary ($dT/dt = 0$) and determined by the balance equations
\begin{align}
 \nu  \Delta T &= \nu^b \Delta T_\perp, \label{eq:tempdiff_balance}\\
 \nu  \Delta T &= -2\nu^b \Delta T_\parallel, \nonumber
\end{align}
in which $\Delta T = T_\parallel - T_\perp$, $\Delta T_\perp = T_\perp -T_\perp^b$, and $\Delta T_\parallel = T_\parallel -T_\parallel^b$.

By observing these temperature differences in the simulation as a function of $\nu^b$, we can determine the point at which $\Delta T = \Delta T_\perp$ (or, equivalently, $\Delta T = -2 \Delta T_\parallel$) and 
thus $\nu^b=\nu$, identifying the unknown physical isotropization rate with the known artificial heat bath frequency. 

\begin{figure}
\rotatebox{0}{\includegraphics{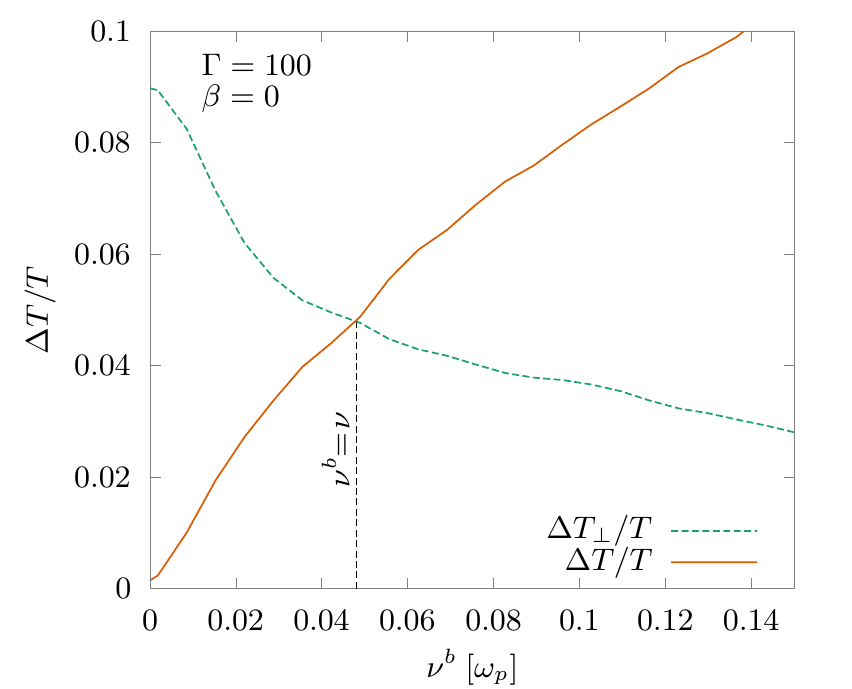} }
\caption{(color online) Relative temperature differences $\Delta T/T$ and $\Delta T_\perp/T$ (see text, $T$ is the equilibrium temperature) as a function of heat bath coupling frequency $\nu^b$. When 
both temperature differences coincide, the heat bath coupling frequency equals the isotropization rate [Eq.~\eqref{eq:tempdiff_balance}].
}  
\label{fig:isotropization_scheme} 
\end{figure}

An exemplary measurement is shown in Fig.~\ref{fig:isotropization_scheme} for \mbox{$\beta=0$}. As the heat bath coupling frequency is increased, the difference in \Tpar and \Tperp 
grows ($\Delta T$ increases) and \Tperp approaches the bath temperature $T_\perp^b$ ($\Delta T_\perp$ decreases). The bath coupling at which $\Delta T = \Delta T_\perp$ and $\nu^b=\nu$ is marked 
by the vertical line. 

\begin{figure}
\rotatebox{0}{\includegraphics{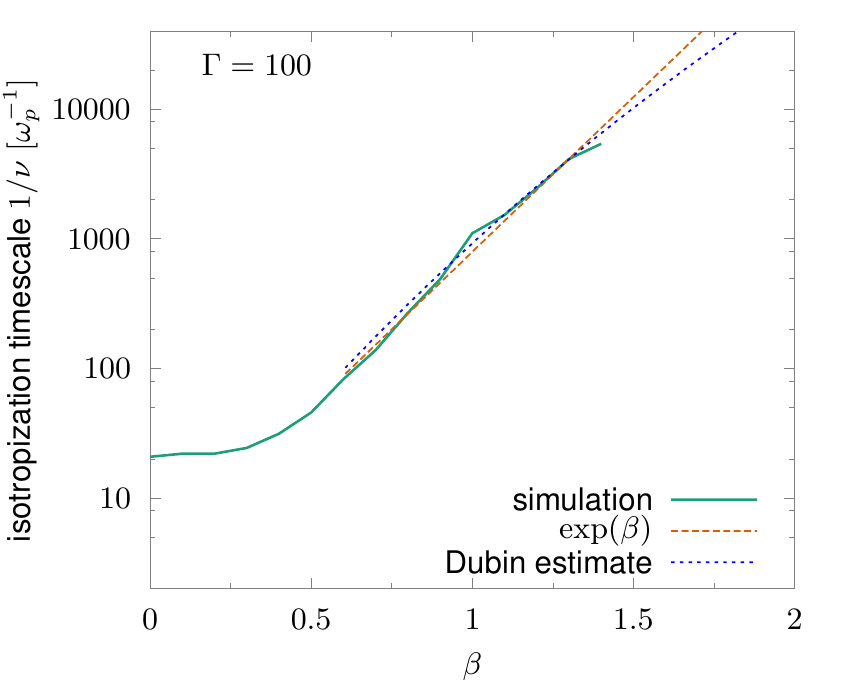} }
\caption{(color online) Isotropization time scale as a function of $\beta$ for $\Gamma=100$. The straight lines shows an exponential extrapolation and the Dubin estimate (see text).  }  
\label{fig:isotropization_g100b} 
\end{figure}

This procedure is repeated for systems with varying magnetic field strength to obtain the isotropization timescale $1/\nu$ as a function of $\beta$, Fig.~\ref{fig:isotropization_g100b}. 
Up to $\beta\approx 0.5$, the isotropization timescale remains approximately constant, which is in line with previous investigations of magnetic field effects~\cite{Ott2013a, Ott2014, Ott2011c}. 
Upon further increase of the magnetic field, the timescale on which isotropization occurs increases drastically. We find that the growth is approximately exponential in $\beta$ 
(recall that $\beta\sim B$), cf. the red (upper) dotted line in Fig.~\ref{fig:isotropization_g100b}. We also show a comparison with the estimate of Dubin~\cite{Dubin2008} 
for the isotropization, $\nu\sim \exp[-\beta\ln(\eta\Gamma/\beta)]/\beta$ with $\eta$ being a constant of order unity, blue dotted line in Fig.~\ref{fig:isotropization_g100b}. Evidently, the data
do not allow to distinguish between the (slightly slower) Dubin estimate and an exponential growth~\footnote{We note that we have treated $\Gamma$ as a free parameter in 
fitting the Dubin estimate to our simulation result to reflect the screened interaction in the simulations. The exact results are dependent on the fitting range, chosen 
here to be $\beta=0.5\dots 1.3$. }.

Overall, our simulations confirm the existence of the many-particle adiabatic invariant predicted by theory and show that the suppression of isotropization is 
approximately an exponential function of the magnetic field strength. 

\begin{figure}
\includegraphics{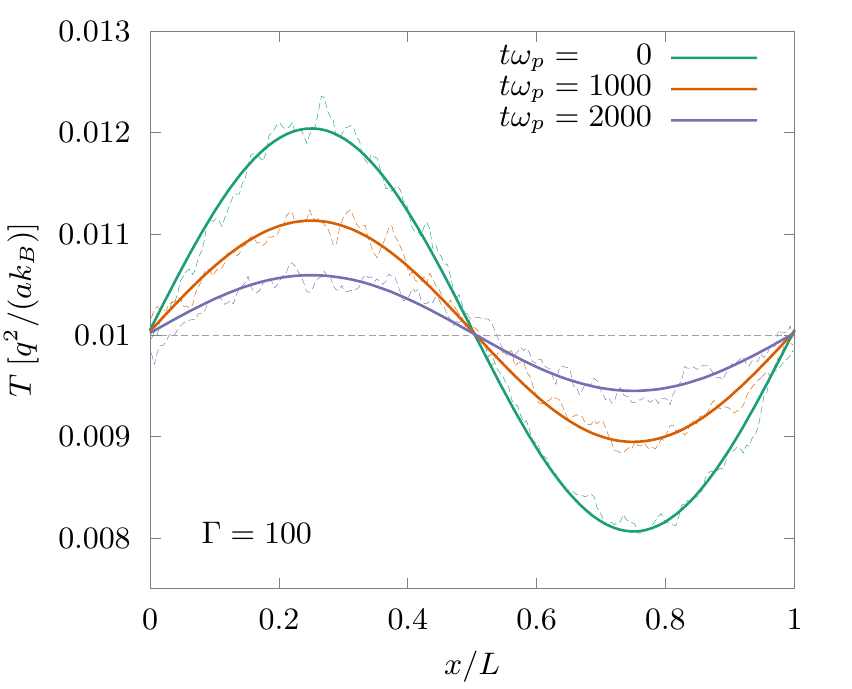} 
\caption{(color online) Temperature profile along the x-axis at the start of the relaxation ($t\omega_p=0$) and at two later times. A sinusoidal fit (solid lines) through the simulation data (dashed lines) is shown 
at each time instant. }
\label{fig:temp_profile_b0} 
\end{figure}

\begin{figure}
\includegraphics{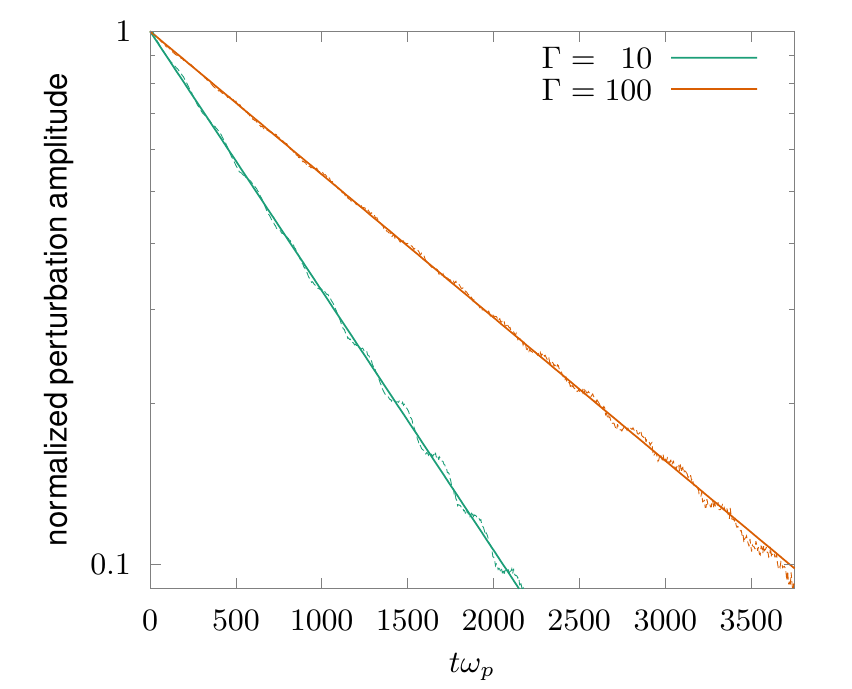} 
\caption{(color online) Time evolution of the normalized temperature perturbation amplitude, cf. Fig.~\ref{fig:temp_profile_b0}, at two different coupling strengths. The solid lines indicate an exponentially decaying fit to the simulation data.  }
\label{fig:amplitude_b0} 
\end{figure}

\section{Temperature Relaxation in Unmagnetized plasmas}
\label{sec:temprelax}
\subsection{Thermal relaxation time scale}
In this section, we consider the relaxation of a spatial temperature perturbation in an unmagnetized plasma. The perturbation 
occurs along a selected axis but encompasses all temperature components, i.e., it is isotropic. 
We follow the prescription of Donk\'o and Ny\'iri~\cite{Donko1998} in creating the temperature perturbation by imposing 
a sinusoidal temperature profile on the plasma. This is advantageous since it results in no net energy change and is trivially 
compatible with the periodic boundary conditions. 

Since the structural relaxation of a magnetized SCP is known to be exponentially slow in the magnetic field strength~\cite{Ott2013a}, we do not create
the sinusoidal temperature profile by a sudden temperature quench (as in Ref.~\cite{Donko1998}) but by coupling the system 
to a Langevin heat bath with the correct temperature profile during the equilibration stage of the simulation ($\nu^b/\omega_p = 0.5$). 

Choosing the $x$-axis as the direction of the perturbation, the temperature profile in Eq.~\eqref{eq:langevin2} is therefore
\begin{align}
T(x)= T_0 + K_0 \sin(kx). \label{eq:sinetemp}
\end{align}
where $k=2\pi/L$, $T_0=q^2/(ak_B\Gamma)$, and $K_0/T_0=0.2$, i.e., the perturbation has a 20\% amplitude.

After the system is prepared with the above temperature profile, the heat bath coupling 
is removed and the system is propagated without further heating in the microcanonic ensemble ($\nu^b/\omega_p=0$). 

The relaxation of the perturbed plasma toward homogeneity is governed by the one-dimensional heat equation, 
\begin{align}
 \frac{\partial T}{\partial t} = \alpha \frac{\partial^2 T}{\partial x^2},\label{eq:heateq}
\end{align}
where $\alpha$ is the thermal diffusivity. 
The solution to \eqref{eq:heateq} for the initial condition \eqref{eq:sinetemp} with periodic boundary conditions is
\begin{align}
 T(x, t) &= T_0 +K(x,t),\\
 K(x,t) &= K_0 \sin(k x) \exp(-t/{\tau^\textrm{th}})\label{eq:decay_sine}
\end{align}
in which 
\begin{align}
\tau^\textrm{th}=\frac{1}{k^2\alpha} \label{eq:timediff}
\end{align} 
is the characteristic temperature homogenization time scale.

\begin{figure}
\rotatebox{90}{\includegraphics{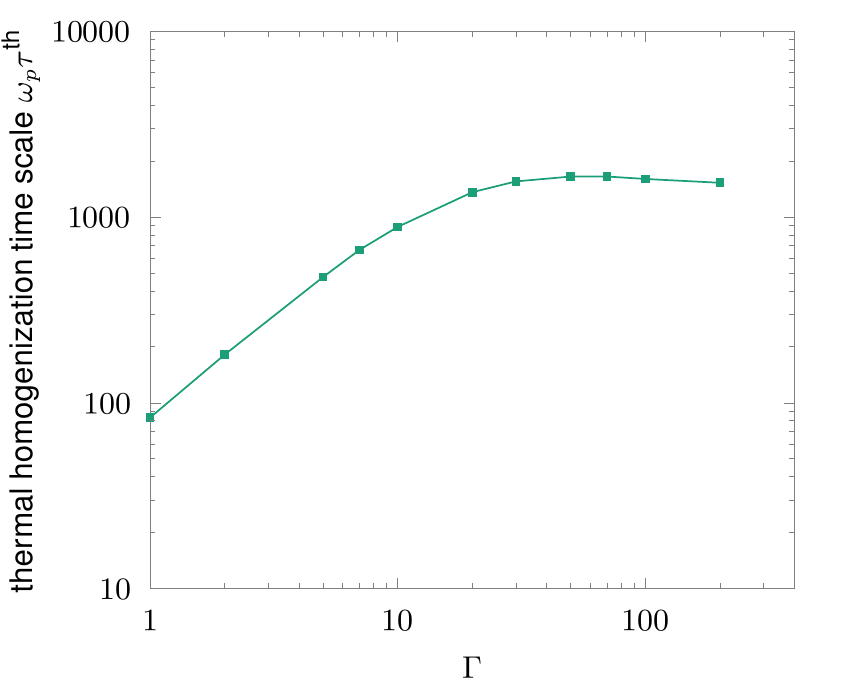} }
\caption{(color online) Time scale of temperature relaxation $\tau^\textrm{th}$ as a function of coupling strength $\Gamma$. }
\label{fig:thermal_timescale_b0} 
\end{figure}

Figure~\ref{fig:temp_profile_b0} shows the temperature profile directly after the heat bath coupling is removed ($t\omega_p=0$) and 
at two later times. Evidently, the sinusoidal shape is conserved while the amplitude diminishes as expected from Eq.~\eqref{eq:decay_sine}. 
The amplitude of the temperature perturbation is shown in Fig.~\ref{fig:amplitude_b0} as a function of time for two values of $\Gamma$, 
confirming the predicted exponential decay. From the measurement of the amplitude decay we extract the homogenization time scale~\eqref{eq:timediff} 
as a function of $\Gamma$, Fig.~\eqref{fig:thermal_timescale_b0}. 

The temperature homogenization time scale shows the well-known non-monotonic dependence on the coupling strength $\Gamma$ which is related to the different 
modes of thermal diffusion~\cite{Ott2015a,Ott2016}. Comparing the isotropization time scale at $\beta=0$ and $\Gamma=100$ (Fig.~\ref{fig:isotropization_g100b}) 
with the corresponding thermal relaxation time scale (Fig.~\ref{fig:thermal_timescale_b0}), it is clear that the former is much shorter and any anisotropy 
in the thermal relaxation process will be quickly removed by collisional isotropization in the unmagnetized plasma. 

\subsection{Thermal conductivity}

To make contact with equilibrium simulations for the thermal conductivity~\cite{Salin2002,Donko2004,Donko2009,Ott2015a}, 
we calculate the thermal conductivity $\lambda$ as it occurs in Fourier's law, 
\begin{align}
 \vec j = -\lambda \nabla T,
\end{align}
where $\vec j $ is the heat flux. The thermal diffusivity $\alpha$ is related to the thermal conductivity $\lambda$ via 
\begin{align}
 \lambda  =\rho c \alpha = \frac{\rho c}{ k^2\tau^\textrm{th}}, \label{eq:diffcond}
\end{align}
where $\rho$ is the mass density and $c$ the specific heat capacity. In the following, we identify $c$ with the 
specific heat capacity at constant volume $c_v$. Since, for the system parameters
at hand, the difference between $c_v$ and $c_p$ (the specific heat capacity at constant pressure) is less than $1\%$~\cite{Khrapak2015a, Khrapak2015}, this 
approximation does not introduce a large error. In addition, the density modulation in the system due to temperature gradient is also less than $1\%$, 
giving further support to a constant volume description. 

An expression for the specific heat $c_v$ of a Yukawa OCP is obtained from the total internal energy $U$, 
\begin{align}
 U &= Nk_BT \bigg[\frac{3}{2}  + u_\textrm{ex}(\Gamma) \bigg],\\
 Mc_v &=\bigg(\frac{\partial U}{\partial T}\bigg)_V =Nk_B \bigg[\frac{3}{2} +  u_\textrm{ex}(\Gamma) - \Gamma\frac{\partial u_\textrm{ex}(\Gamma)}{\partial \Gamma}\bigg], \label{eq:cv}
\end{align}
where $u_\textrm{ex}$ is the reduced excess energy and $M$ is the system mass. 

\begin{figure}
\rotatebox{0}{\includegraphics{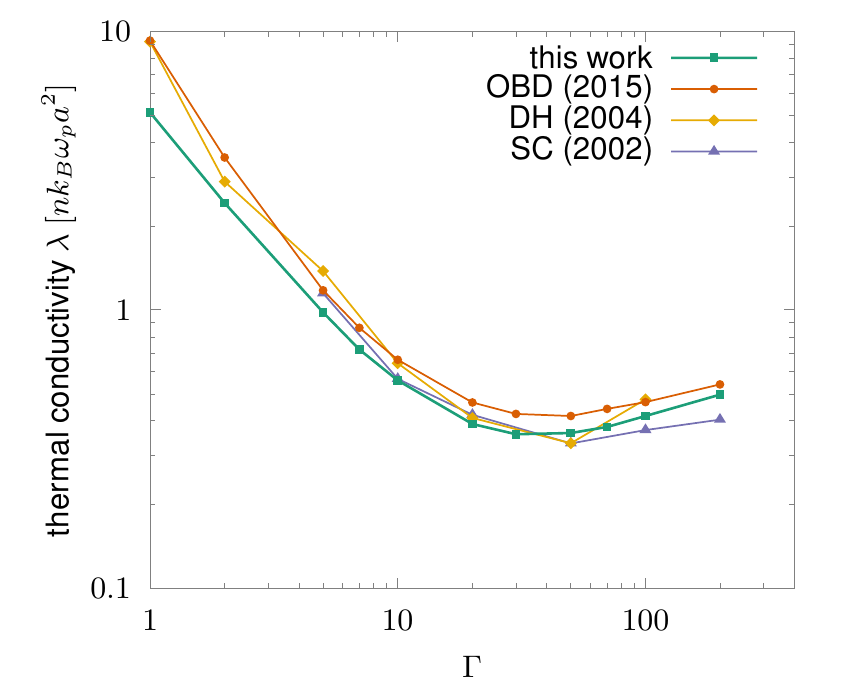} }
\caption{(color online) Thermal conductivity calculated from temperature relaxation time scale. For comparison, results from Refs.~\cite{Salin2002} (SC) and~\cite{Ott2015a} (OBD)
based on the Green-Kubo relations and from Ref.~\cite{Donko2004} (DH) based on a non-equilibrium method are shown.}
\label{fig:heat_conductivity_b0} 
\end{figure}

\begin{figure}
\includegraphics{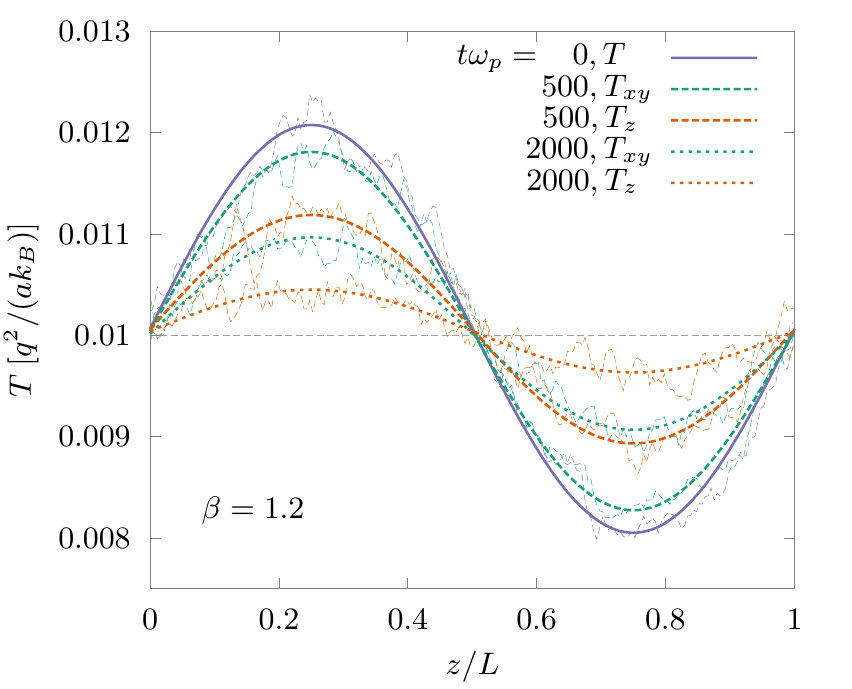} 
\caption{(color online) Profiles of the field-parallel and field-perpendicular temperatures in a magnetized system at two different times. 
The temperature gradient is oriented along the magnetic field.  A sinusoidal fit through the data is shown 
at each time instant. }
\label{fig:temp_profile_b12} 
\end{figure}

For $u_\textrm{ex}(\Gamma)$, Khrapak and Thomas~\cite{Khrapak2015} propose the following expression, 
\begin{align} 
 u_\textrm{ex}(\Gamma) = \delta (\Gamma/\Gamma_m)^{2/5} + \epsilon + C_f \Gamma, \label{eq:khrapakfit}
\end{align}
with $\delta=3.2$ and $\epsilon = -0.1$~\cite{Khrapak2015}. Here, $\Gamma_m=440$ and $C_f=0.1042$ are the melting point~\cite{Hamaguchi1997} and fluid Madelung constant~\cite{Rosenfeld2000} for $\kappa=2$, respectively. 

From this, the thermal conductivity can be calculated and compared with earlier results~\cite{Salin2002,Donko2004,Ott2015a}, see Fig.~\ref{fig:heat_conductivity_b0}. 
We find convincing qualitative agreement of the functional dependence although the thermal conductivity is consistently underestimated by about 20\% compared to Ref.~\cite{Ott2015a}. Sources for 
this deviation include the comparably large deviation from equilibrium in the simulations at hand as well as uncertainties in the available data for the 
total internal energy. Comparable deviations were found in Ref.~\cite{Donko1998} for the Coulomb OCP.

\section{Temperature Relaxation in magnetized plasmas}
\label{sec:magtemprelax}
We now turn to the temperature relaxation in magnetized plasmas. We apply the same procedure of coupling the system 
to a sinusoidally modulated heat bath as before, after which the system relaxation is monitored in the microcanonic ensemble. 
Choosing the $z$-axis as the direction of the magnetic field, we consider a temperature perturbation along the field lines, $\nabla T\parallel \vec B$. 

Again, an exemplary result is shown in Fig.~\ref{fig:temp_profile_b12} where $\beta=1.2$. 
The locally isotropic temperature perturbation at $t\omega_p=0$ is identical to the unmagnetized case considered before. Since 
the temperature relaxation now occurs in the presence of a magnetic field, this process is modified. As Fig.~\ref{fig:temp_profile_b12} 
shows, the time evolution of the field -parallel and the cross-field temperatures $\Tpar=T_z$ and $\Tperp=T_{xy}$ is decoupled 
and a temperature anisotropy develops in the plasma. We stress again that this is not caused by an anisotropic heating or cooling mechanism, but 
solely by the anisotropic thermal diffusivity of the system. 

To model the temperature relaxation in this plasma, we generalize Eq.~\eqref{eq:heateq} to include the isotropization process 
via the isotropization rate $\nu$, 
\begin{align}
 \frac{d\Tperp}{dt} &= \alpha^z_\perp \frac{d^2\Tperp}{dz^2} - \nu (\Tperp-\Tpar),\label{eq:tempevperp}\\
 \frac{d\Tpar}{dt} &= \alpha^z_\parallel \frac{d^2\Tpar}{dz^2} - 2 \nu (\Tpar-\Tperp).\label{eq:tempevpar}
\end{align}
Note that the sum of Eq.~\eqref{eq:tempevperp} multiplied by two and Eq.~\eqref{eq:tempevpar} evaluates to the heat equation~\eqref{eq:heateq}. 
The subscripts of $\alpha^z_{\perp/\parallel}$ indicate the temperature component relative to $z$ (the direction of the temperature gradient). 
This should not be identified with the thermal diffusivity of a plasma parallel and 
perpendicular to a magnetic field which was investigated in Ref.~\cite{Ott2015a}. For notational clarity, we drop the superscript $z$ in the following. 

Writing the temperatures as sums of the equilibrium temperature $T$ and modulations $\hat T_\perp$ and $\hat T_\parallel$,
\begin{align}
 T_\perp &= T + \hat T_\perp(z,t),\\
 T_\parallel &= T + \hat T_\parallel(z,t) 
\end{align}
and using the ansatz
\begin{align}
 \hat T_\perp(t, z) &= K_\perp(t) \sin(kz),\\
 \hat T_\parallel(t, z) &= K_\parallel(t) \sin(kz), 
\end{align}
where $K_\parallel$ and $K_\perp$ are the time-dependent amplitudes, 
Eqs.~\eqref{eq:tempevperp} and \eqref{eq:tempevpar} reduce to the ordinary differential equations
\begin{align}
 \frac{dK_\perp}{dt} &= -\alpha_\perp k^2 K_\perp(t) - \nu (K_\perp - K_\parallel),\\
 \frac{dK_\parallel}{dt} &= -\alpha_\parallel k^2 K_\parallel - 2 \nu (K_\parallel-K_\perp)
\end{align}
with solutions
\vskip1cm
\begin{widetext}
\begin{align}
K_\parallel(t)/K_0 &= \frac{e^{-\frac{1}{2} t \left(k^2 (\alpha_\parallel+\alpha_\perp)+3 \nu \right)} \left(\sinh \left(\frac{g
   t}{2}\right) \left(k^2 (\alpha_\perp-\alpha_\parallel)+3 \nu \right)+g \cosh \left(\frac{g t}{2}\right)\right)}{g}\label{eq:apart}\\
K_\perp(t)/K_0 &= \frac{e^{-\frac{1}{2} t \left(k^2 (\alpha_\parallel+\alpha_\perp)+3 \nu \right)} \left(\sinh \left(\frac{g
   t}{2}\right) \left(k^2 (\alpha_\parallel-\alpha_\perp)+3 \nu \right)+g \cosh \left(\frac{g t}{2}\right)\right)}{g}\label{eq:aperpt}
\end{align}
\end{widetext}
where 
\begin{align}
g=\sqrt{k^4 (\alpha_\parallel-\alpha_\perp)^2+2 k^2 \nu (\alpha_\parallel-\alpha_\perp)+9 \nu^2}
\end{align}
and $K_{\parallel/\perp}(t=0)=K_0$. 

Note that for the special case $\alpha_\parallel=\alpha_\perp$ as well as in the fast isotropization limit $\nu\gg k^2(\alpha_\parallel-\alpha_\perp)$, Eqs.~\eqref{eq:apart} and \eqref{eq:aperpt} reduce to 
\begin{align}
K_\parallel(t) &= K_\perp(t) = K_0 \exp(-k^2(\alpha_\parallel+\alpha_\perp)/2 \cdot t),
\end{align}
which is consistent with Eq.~\eqref{eq:decay_sine}. 

In the following, we first discuss the general solution types of Eqs.~\eqref{eq:apart} and \eqref{eq:aperpt} which are then compared with the simulation data. 

\begin{figure}
\includegraphics{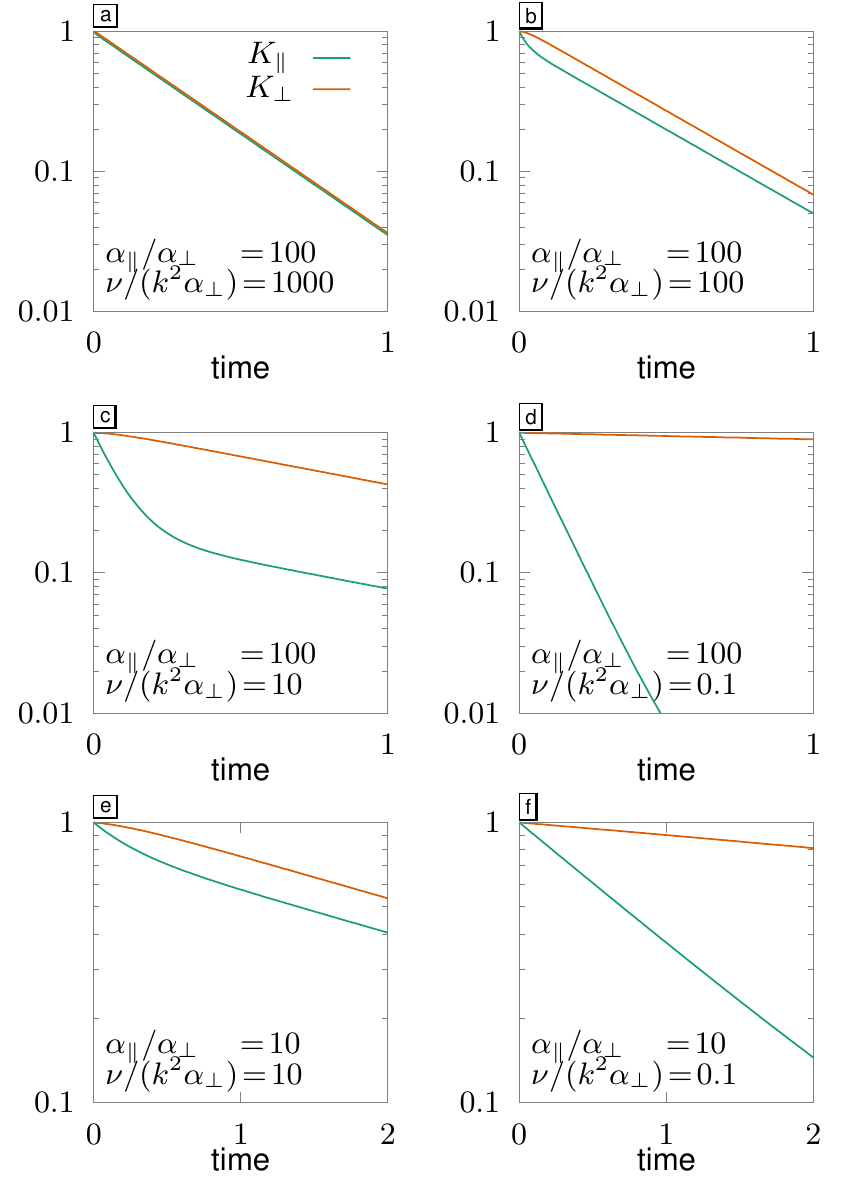} 
\caption{(color online) Solution (normalized) of Eqs.~\eqref{eq:apart} and \eqref{eq:aperpt} for different ratios of the thermal diffusivities and isotropization ratio.}
\label{fig:theoretical_curves} 
\end{figure}

The solutions~\eqref{eq:apart} and \eqref{eq:aperpt} depend on (the ratio of) three parameters, $k^2\alpha_\perp$, $k^2\alpha_\parallel$, and $\nu$. 
In Fig.~\ref{fig:theoretical_curves}, we explore several combinations of these three parameters which showcase the different solution types. Subfigures (a)-(d) 
depict a case in which the thermal diffusivities $\alpha_{\perp}$ and $\alpha_\parallel$ differ strongly, e.g., $\apar/\aperp=100$. 
The temperature isotropy of the system now crucially depends on the isotropization rate: If $\nu\gg k^2\alpha$ (Subfig. a), no anisotropy occurs since any 
difference between \Tpar and \Tperp is quickly removed. If $\nu$ is comparable to (Subfig. b) or smaller than $k^2\alpha$ (Subfig. c), an anisotropy develops which 
grows with decreasing $\nu$. Note that the temperature associated with greater thermal diffusivity initially drops sharply but then decays 
with the same exponential rate as the higher temperature. The temperature ratio $\Tperp/\Tpar$ thus remains constant. 

If isotropization is further reduced (Subfig. d), the initial drop in $\Tpar$ is more rapid. In this situation, \Tpar quickly becomes spatially homogeneous while 
\Tperp remains sinusoidally modulated for increasingly long times. 

In Subfigs. (e) and (f), we investigate the solutions at $\apar/\aperp=10$ and find qualitatively 
similar behavior (note the different data ranges in these subfigures). 

\begin{figure}
\includegraphics{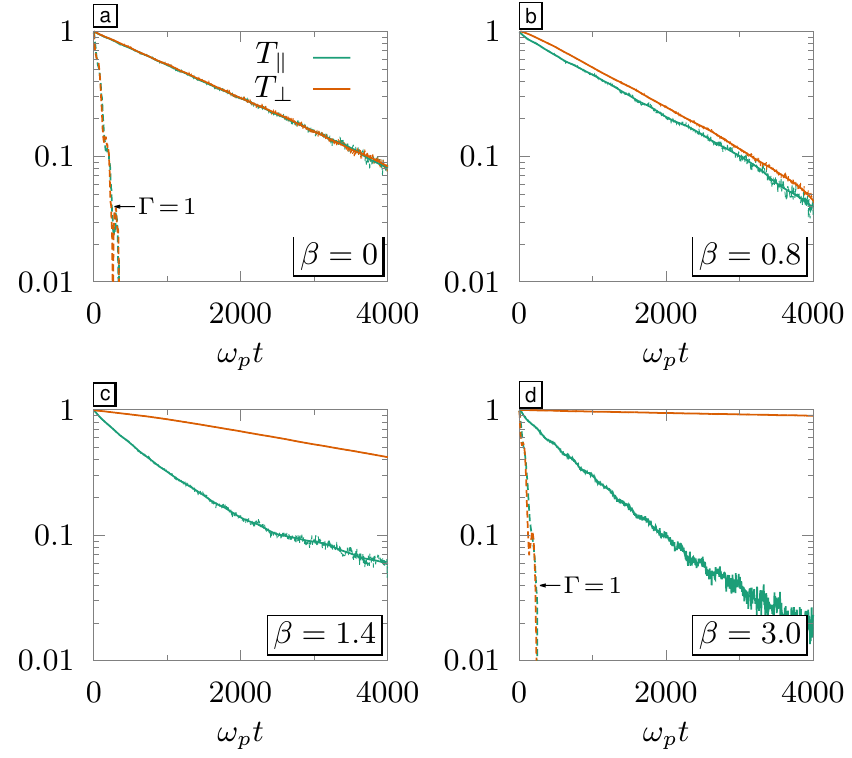} 
\caption{(color online) Temporal dependence of sinusoidal amplitude (normalized) for different magnetic fields, $\nabla T\parallel \hat{\vec e}_z$, and $\Gamma=100$. 
Compare with Fig.~\ref{fig:theoretical_curves}(a-d). The dashed lines in subgraphs (a) and (d) show data for $\Gamma=1$. }
\label{fig:simulation_curves} 
\end{figure}

Having considered the theoretical solutions, we now turn to the simulation results. As before, we extract the amplitude of the sinusoidal modulation as a function of time, 
separately for \Tpar and \Tperp (see Fig.~\ref{fig:temp_profile_b12}). Typical results for different 
values of $\beta$ are shown in Fig.~\ref{fig:simulation_curves}. Comparing with the theoretical curves, Fig.~\ref{fig:theoretical_curves}(a)-(d), 
we find convincing qualitative agreement between theory and simulation.

\begin{figure}
\includegraphics{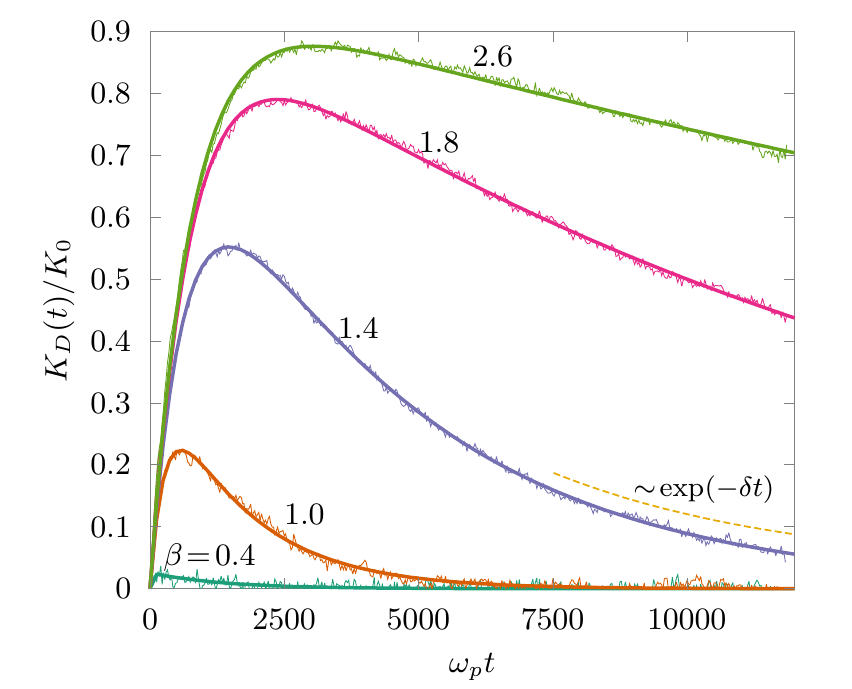} 
\caption{(color online) Anisotropy $K_D(t) = K_\perp(t) - K_\parallel(t)$  (simulation results, thin lines) and 
best fit to theoretical curves (strong lines) for different values of $\beta$. The broken line segment follows an exponential decay $\exp(-\delta t)$ as a guide for 
the eye for $\beta=1.4$. }
\label{fig:KD_wfit} 
\end{figure}

To quantify this agreement, we now consider, as a measure of temperature anisotropy in the plasma, the difference in the temperature modulation amplitudes, 
\begin{align} 
K_D(t) = K_\perp(t) - K_\parallel(t).  \label{eq:kd}
\end{align}

If an anisotropy develops, $K_D(t)$ will be a peaked function with limits $K_D(0)=K_D(\infty) = 0$. From Eqs.~\eqref{eq:apart}-\eqref{eq:aperpt}, the 
long-time asymptotic corresponds to an exponential decay with decay constant $\delta = (k^2(\alpha_\perp+\alpha_\parallel)+3\nu + g)/2$, i.e., 
$K_D(t) \sim \exp(-\delta t)$ for large $t$. 

In Fig.~\ref{fig:KD_wfit}, 
simulation results for $K_D(t)$ are shown together with best fits to the theoretical results. Evidently, the temperature anisotropy develops quickly (for sufficiently high 
magnetic fields) as the field-parallel temperature perturbation is removed by the thermal diffusivity ($K_\parallel \rightarrow 0$). The inhibited isotropization and 
the decreased thermal diffusivity in the cross-field temperature component leads to increasingly pronounced and increasingly long-lived temperature anisotropies 
which decay exponentially with time. 

\begin{figure}
\includegraphics{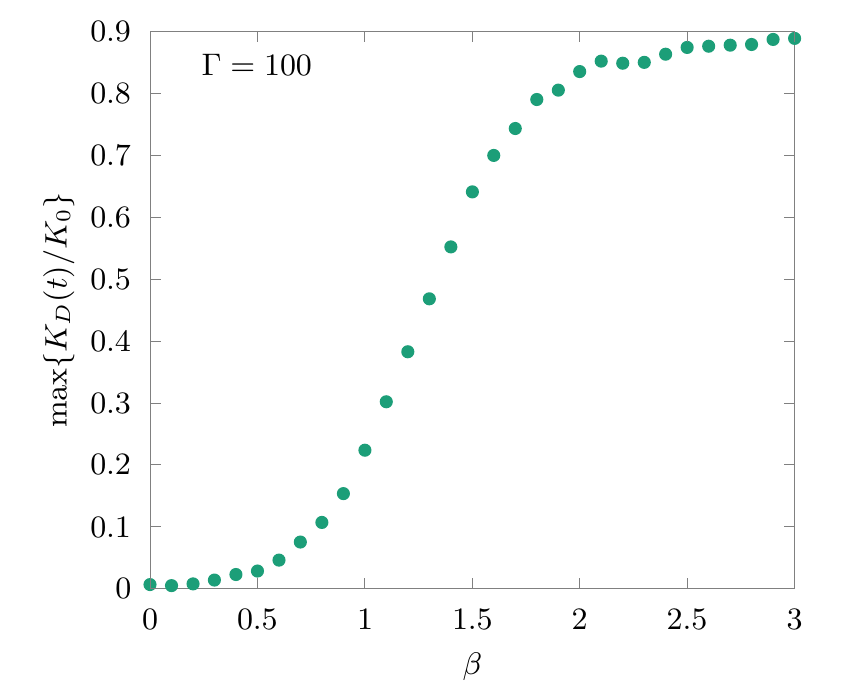} 
\caption{(color online) Maximum value of $K_D(t) = K_\perp(t) - K_\parallel(t)$ relative to $K_0$ as a function of $\beta$ (simulation data). }
\label{fig:max_anisotropy_z} 
\end{figure}

\begin{figure}
\includegraphics{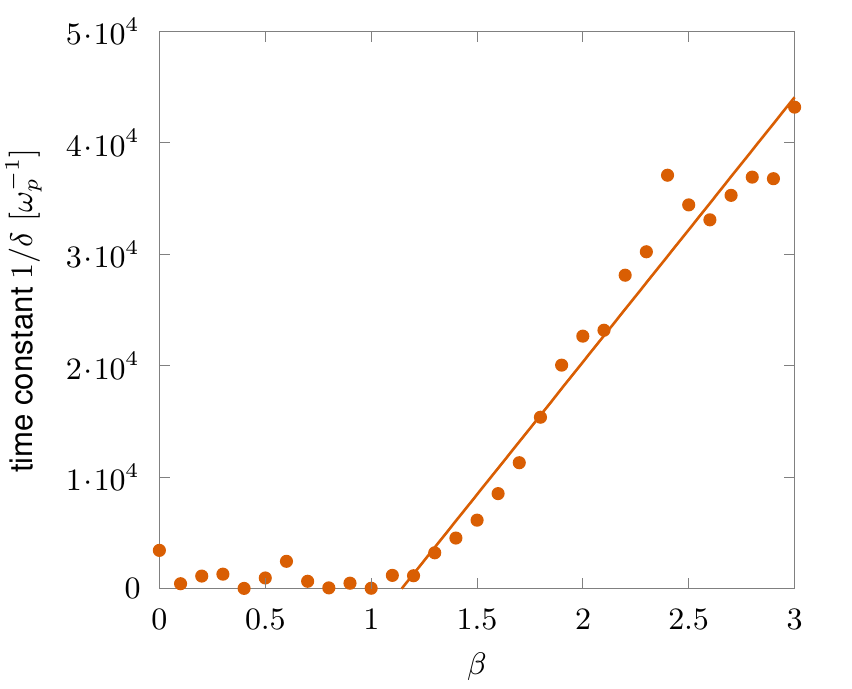} 
\caption{(color online) Time constant of the exponential decay of $K_D(t)\sim\exp(-\delta t)$ at long times (simulation data). }
\label{fig:anisotropy_time_z} 
\end{figure}

To quantify these finding, we consider the peak value of $K_D(t)$ and the asymptotic decay constant, Figs.~\ref{fig:max_anisotropy_z} and~\ref{fig:anisotropy_time_z}. 
As Fig.~\ref{fig:max_anisotropy_z} shows, a sizable anisotropy develops at a critical magnetic field strength of $\beta_\textrm{c}\approx 1.1$ and rapidly 
increases with increasing $\beta$ (it is limited to a maximum $K_D(t)/K_0=1$ by construction). This is accompanied by a large enhancement of the 
anisotropy longevity which is reflected in an increase of inverse decay constant $1/\delta$ (Fig.~\ref{fig:anisotropy_time_z}). This enhancement 
is approximately linear in $\beta$ for $\beta>\beta_\textrm{c}$. 

\begin{figure}
\includegraphics{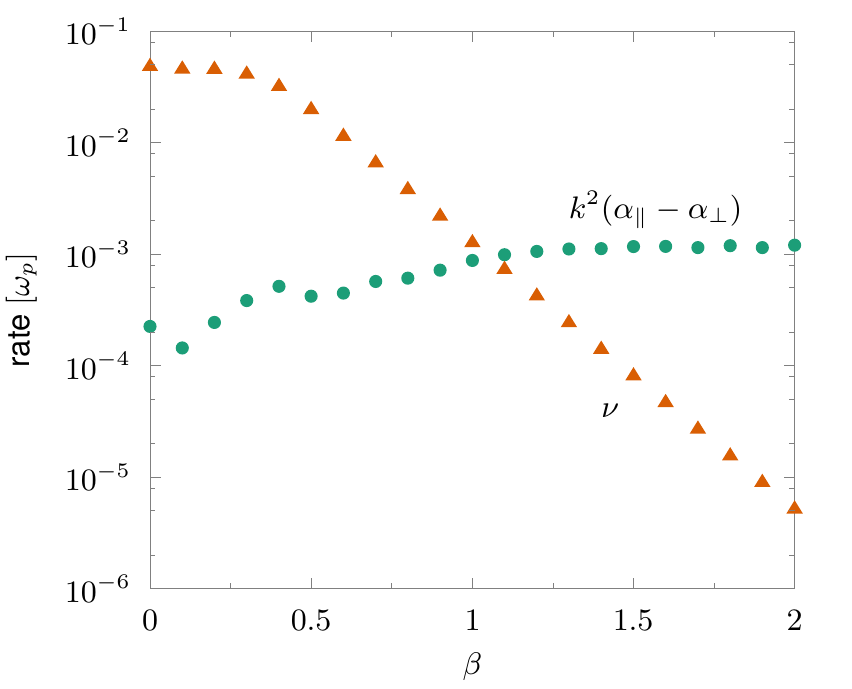} 
\caption{(color online) Isotropization rate $\nu$ and effective anisotropization rate $k^2(\alpha_\parallel-\alpha_\perp)$ as a function of beta [data from fit to $K_D(t)$]. }
\label{fig:iso_vs_aniso} 
\end{figure}

To complete this analysis, we use the functional form~\eqref{eq:kd} of~$K_D(t)$ to extract the effective anisotropization rate $k^2(\alpha_\parallel - \alpha_\perp)$ 
from the simulation data. To increase accuracy, we utilize the earlier results for $\nu(\beta)$ as input, treating only the thermal diffusivities as 
free parameters during the fit. This allows for a direct comparison of the two competing processes shown in Fig.~\ref{fig:iso_vs_aniso}. 
At small magnetic fields, the isotropization process due to collisions is much faster than the difference in the thermal diffusivities and any 
anisotropy is quickly removed. As $\beta$ increases, $\nu$ is drastically reduced as discussed before, whereas the anisotropization rate increases. 
This latter effect is due to the enhancement of field-parallel and the concomitant reduction in cross-field collisional thermal conduction~\cite{Ott2015a, Ott2016}. 
At $\beta_\textrm{c}\approx 1.1$, the difference in the thermal diffusivities is larger than the isotropization and the observed anisotropy develops.

\subsection{Discussion}
We now discuss the physical mechanism for the generation of the observed anisotropy. Clearly, one of the underlying causes is the suppression of isotropization 
by the magnetic field. Since only collisions and, by extension, particle-wave interactions, can mediate isotropization, the magnetic field 
must reduce the frequency or isotropization effectiveness of these collisions. In strong magnetic fields, the particles move on tight helical 
trajectories along the field lines and the collisional energy transfer is therefore reduced across the field lines~\cite{Ott2015a}, which is the main contribution 
to the reduced isotropization. 

The second cause of the anisotropy generation is the existence of a difference in the thermal diffusivities $\alpha_\parallel$ and $\alpha_\perp$ associated with
the different temperature components, which provides the actual source for the anisotropy. 
In a simple particle-based picture, thermal diffusion is the result of two different processes: 
One is the direct movement of particles from hotter regions into colder ones and vice-versa. Since kinetic and potential energy are associated with each particle, the result is 
a net energy transfer. This bodily movement of particles is the main contribution to thermal diffusion in weakly correlated plasma. 
The second mechanism are collisions between particles which result in a momentum exchange and energy transfer. 
This process dominates in strongly correlated plasmas ($\Gamma\gtrsim 30$)~\cite{Ott2015a}. 

\begin{figure}
\rotatebox{0}{\includegraphics{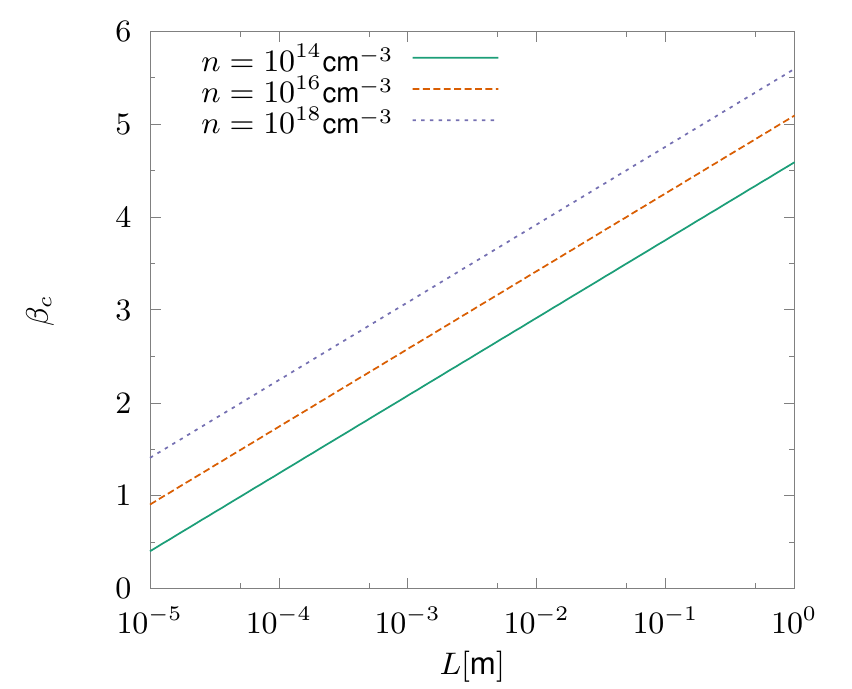} }
\caption{(color online) Estimation of critical magnetic field strengths as a function of the perturbation length for various plasma densities. }
\label{fig:criticalbL} 
\end{figure}

The first process (particle migration) is isotropic in the sense that field-parallel and cross-field momentum are both transferred simultaneously. The second 
process (collisions), however, is anisotropic as only momentum along the collisional axis is transferred. 
Since the magnetic field causes an imbalance in the frequency and efficacy of the field-parallel and cross-field collisions, 
the total collisional contribution toward $\alpha_\parallel$ is different from that toward $\alpha_\perp$. 

Therefore, in plasmas in which collisional processes are the primary cause of thermal diffusion (i.e., in plasmas with $\Gamma\gtrsim 30$), the value of 
$\alpha_\parallel - \alpha_\perp$ is finite and positive. For plasmas in which the isotropization is fast enough to mask this difference (cf. Fig.~\ref{fig:iso_vs_aniso}), 
no anisotropy develops---any difference between \Tpar and \Tperp is quickly removed. However, once the isotropization frequency is sufficiently lowered by 
the magnetic field to become less than $k^2(\alpha_\parallel - \alpha_\perp)$, a difference in \Tpar and \Tperp will be observed. 

To verify this physical mechanism, we have carried out simulations for a system in which thermal diffusion is dominated by bodily movement 
and collisional processes play no major role ($\Gamma=1$). As expected, no appreciable anisotropy is observed even for the largest magnetic field strength considered ($\beta=3$), 
see Fig.~\ref{fig:simulation_curves}(a) and~(d). 

\section{Summary}
\label{sec:summary}

In summary in this work, we have addressed three main issues: First, we have investigated the isotropization process in SCPs both in the absence and in the presence 
of external magnetic fields. Using a balance-equation approach, we were able to isolate the measurement of the isotropization from confounding structural rearrangement 
and to extract accurate data for the isotropization rate. In agreement with theoretical predictions by O'Neil and Hjorth~\cite{O’Neil1983,O’Neil1985}, we have found 
that a magnetic field strongly restricts the ability of the plasma to equalize temperature anisotropies. The functional dependence of this suppression was found to 
be approximately exponential in the magnetic field strength. 

Secondly, we have demonstrated that a non-equilibrium molecular dynamics simulation approach based on a sinusoidal temperature perturbation~\cite{Donko1998}
is able to reproduce earlier results for the thermal conductivity of unmagnetized plasmas~\cite{Ott2015a} and can be used to extract the physical timescale of the 
thermal diffusion process. 

Finally, we have investigated the effect of a magnetic field on a spatial temperature perturbation in a SCP when the temperature gradient 
is along the field lines. Here, the crucial factor is that in SCPs, thermal diffusion is primarily mediated by collisions which only transfer momentum along the axis of collision. 
If no efficient isotropization process is counteracting this process, the result is a temperature anisotropy as we have observed. This anisotropy occurs beyond a critical 
value of the magnetic field after which it becomes increasingly more prominent and long-lived. 

We have modeled these effects by extending the heat equation to include isotropization. Using the earlier data for the isotropization timescale, we have shown that the 
onset of the anisotropy coincides with the point at which the isotropization rate becomes slower than the effective anisotropization rate, at $\beta_c\approx 1.1$. 

Under the conditions of strong coupling, the temperature anisotropy thus develops intrinsically from the isotropic temperature perturbation. This is unlike the more familiar 
situation of temperature anisotropies in weakly coupled plasmas which are driven by anisotropic energy sources or drains. We expect the effect to be strongest for 
systems in which collisional heat transfer dominates and no other heat dissipation mechanisms (e.g., thermal diffusion by other plasma species) 
interact efficiently with the strongly coupled particles. The maximum temperature difference of field-parallel and cross-field temperature components is naturally 
given by the size of the initial temperature perturbation. 

Let us briefly comment on the importance of the spatial form of the perturbation. Although we have used a sinusoidal modulation in our simulations, this is 
not a necessary restriction and the effect also occurs if truly localized heating is used, e.g., of a Gaussian shape~\footnote{Note, however, that a Gaussian 
perturbation does not fulfill the periodic boundary conditions used in the simulation.}. Since Eq.~\eqref{eq:sinetemp} is the 
fundamental solution for the heat equation with periodic boundary conditions, all other perturbations can be expressed as a linear combination of this solution. 

Lastly, we consider the relevant plasma parameters at which this effect occurs. The timescale of homogenization $\tau^\textrm{th}$ scales quadratically in 
the linear size of the perturbation,  $\tau^\textrm{th}\sim L^2$ [Eq.~\eqref{eq:timediff}] while the isotropization timescale is exponential in $\beta$ 
or very near so (Fig.~\ref{fig:isotropization_g100b}). One can thus obtain the critical value of $\beta$ necessary for the emergence of a temperature anisotropy 
as a function of the perturbation length. The results are shown in Fig.~\ref{fig:criticalbL} for three plasma densities. 
Evidently, perturbing the plasma on very large length scales results in an isotropic temperature even at high magnetic fields. On the other hand, small scale perturbations caused 
by localized energy sources can lead to an anisotropy already at small critical magnetic field strengths. 

The influence of a second plasma species remains an open question at this point which will be addressed in future work.

\begin{acknowledgments}
This work is supported by the DFG via SFB-TR 24 project A7, grant shp00014
for CPU time at the North-German Supercomputing Alliance HLRN, 
NKFIH Grant K-115805, the J\'anos Bolyai Research Scholarship of the Hungarian Academy of Sciences (PH), 
and NKFIH Grant K-119357 (ZD).
\end{acknowledgments}

\bibliography{anisotropy}

\end{document}